\documentstyle[12pt]{article}
\newcommand{\T}{{\cal T}}
\renewcommand{\H}{{\cal H}}
\newcommand{\D}{{\cal D}}
\newcommand{\F}{{\cal F}}
\newcommand{\la}{\label}
\newcommand{\non}{\nonumber}
\newcommand{\be}{\begin{equation}}
\newcommand{\ee}{\end{equation}}
\newcommand{\ba}{\begin{eqnarray}}
\newcommand{\ea}{\end{eqnarray}}
\newcommand{\bastar}{\begin{eqnarray*}}
\newcommand{\eastar}{\end{eqnarray*}}
\begin{document}

\begin{titlepage}
 
\begin{center}
\vskip 4.0cm
{\bf \large ABELIAN DECOMPOSITION OF \\ \vskip 0.3cm
            Sp(2N) YANG-MILLS THEORY \\}
\end{center}
 
\vskip 2.0cm
 
\begin{center}
{\bf Wang-Chang Su$^{*}$ }  \\
\vskip 0.3cm
{\it Department of Physics \\
     National Chung-Cheng University, Chia-Yi, Taiwan } \\
\end{center}
 
\vskip 0.5cm

{\rm
In the previous paper, we generalized the method of Abelian decomposition to the case of $SO(N)$ Yang-Mills theory. 
This method that was proposed by Faddeev and Niemi introduces a set of variables for describing the infrared limit of a Yang-Mills theory. 
Here, we extend the decomposition method further to the general case of four-dimensional $Sp(2N)$ Yang-Mills theory. 
We find that the $Sp(2N)$ connection decomposes according to irreducible representations of $SO(N)$. 
}

\noindent\vfill
 
\begin{flushleft}
\rule{5.1 in}{.007 in} \\
$^{*}$ \hskip 0.2cm {\small  E-mail: \bf suw@phy.ccu.edu.tw } \\
\end{flushleft}

\end{titlepage}

It is widely believed that the ultraviolet and infrared limits of a Yang-Mills theory should represent different phases. 
In the high-energy limits the theory describes the interaction of massless gluons and can be solved perturbatively thanks to the asymptotic freedom.
At low energies, the method of perturbation fails because the Yang-Mills theory becomes strongly coupled. 
Some nonperturbative techniques have to be developed.
Despite this, we expect that the low energy limits of the theory should exhibit color confinement due to the dual Meissner effect \cite{nambu,thooft}.
Hence, the spectrum of the theory at low energies possesses massive composites of gauge fields such as glueballs. 

Recently, Faddeev and Niemi have conjectured a novel decomposition of the four-dimensional $SU(2)$ Yang-Mills connection in terms of new variables, appropriate for describing the theory in the infrared limits.
At a certain phase, the decomposed theory known as the Skyrme-Faddeev model supports knotlike solutions, which can be regarded as the candidates for glueballs \cite{fad0,fad1}. 
The method of Faddeev-Niemi decomposition has attracted much research attention since it was introduced.
For instances, the method of decomposition is soon generalized to the case of $SU(N)$ Yang-Mills theory in \cite{periwal,fad2}.
It is also generalized for the $SO(N)$ Yang-Mills theory by the author recently \cite{su}.
The connections on the topological aspects between the original $SU(2)$ Yang-Mills theory and the Skyrme-Faddeev model are investigated in \cite{konishi}. 
The derivation on the Skyrme-Faddeev model from the first principles can be found in \cite{cho1,lang,li}.

In the present letter, we extend the Abelian decomposition method further to the general case of the four-dimensional $Sp(2N)$ Yang-Mills theory. 
We build $N$ mutually orthogonal Lie-algebra valued fields $m_i$ in such a way that they describe $2N^2$ independent variables.
Then using these fields $m_i$, we construct the $Sp(2N)$ covariant one-forms that are orthogonal to the fields $m_i$ and determine a basis of roots in the $Sp(2N)$ Lie algebra.
Consequently, the fields $m_i$ and the covariant one-forms are used to decompose the generic $Sp(2N)$ connection. 

The symplectic Lie group $Sp(2N)$ is rank $N$ and its Lie algebra has $N(2N+1)$ generators.
In the defining representation, the generators are represented by $2N \times 2N$ matrices, which can be further reexpressed in tensor products of $2 \times 2$ and $N \times N$ matrices.
We classify these generators into the following three categories,
\ba
\T_{0,\alpha} & = & \frac{1}{\sqrt{2}} \left( \sigma_0 \otimes T_\alpha \right), 
\la{asymm} \\ 
\T_{I,a} & = & \frac{1}{\sqrt{2}} \left( \sigma_I \otimes T_a \right),
\la{osymm} \\
\T_{I,i} & = & \frac{1}{\sqrt{2}} \left( \sigma_I \otimes T_i \right).
\la{dsymm}  
\ea
In the above equations, $\sigma_0$ is the $2 \times 2$ identity matrix and $\sigma_I$ for \( I = 1,2,3 \) are the Pauli matrices.
$T_\alpha$ for $\alpha$ = 1 to $\frac{1}{2}N(N-1)$ are antisymmetric $N \times N$ matrices. 
$T_a$ for $a$ = 1 to $\frac{1}{2}N(N-1)$ are {\it off-diagonal} symmetric $N \times N$ matrices.
$T_i$ for $i$ = 1 to $N$ are {\it diagonal} symmetric $N \times N$ matrices.
Thereinafter, we shall reserve the first few Greek letters \( ( \alpha, \beta, \gamma ) \) for the indices of $N \times N$ antisymmetric matrices.
Similarly, the first few Latin letters \( ( a,b,c ) \) are reserved for the indices of $N \times N$ off-diagonal symmetric matrices, and the middle Latin letters \( ( i,j,k,l,m,n ) \) are for the indices of $N \times N$ diagonal symmetric matrices.

It is not difficult to see that (\ref{asymm}), (\ref{osymm}), and (\ref{dsymm}) give a total number of $N(2N+1)$ generators, in which the $N$ commuting generators in the Cartan subalgebra are
\be 
\H_i \equiv \T_{3,i} = \frac{1}{\sqrt{2}} \left( \sigma_3 \otimes T_i \right).
\la{Hi} 
\ee
These generators are normalized to 
\ba 
{\rm Tr} \left( \T_{\mu,A} \T_{\nu,B} \right) = 
\frac{1}{2} \delta_{\mu \nu} \delta_{A B},
\la{normalization}
\ea
where $\T_{\mu,A}$ $(\T_{\nu,B})$ stands for 
\( \left( \T_{0,\alpha}, \T_{I,a}, \T_{I,i} \right) \).
In this way, a generic $Sp(2N)$ Lie-algebra element $v$ has the expansion in terms of the generators as \( v = v^{0,\alpha} \T_{0,\alpha} + v^{I,a} \T_{I,a} + v^{I,i} \T_{I,i}. \) 
 
As a matter of fact, the $N^2$ $N \times N$ matrices $T_\alpha$, $T_a$, and $T_i$ introduced in (\ref{asymm}), (\ref{osymm}), and (\ref{dsymm}) can be used to generate the defining representation of the $U(N)$ Lie algebra. 
The Cartan subalgebra is constituted by the matrices $T_i$ \( ( i = 1,\dots,N ) \).
Let us denote the defining representation of the $U(N)$ Lie algebra by \( T_A = \left( T_\alpha, T_a, T_i \right) \) with the multiplication law 
\be
T_A T_B = \frac{i}{2} f_{ABC} T_C + \frac{1}{2} d_{ABC} T_C,
\la{fabcdabc}
\ee
where \( A,B,C = 1,2,\dots,N^2 \). 
The structure constants $f_{ABC}$ are completely antisymmetric and real, whereas the coefficients $d_{ABC}$ are the completely symmetric. 
From (\ref{fabcdabc}), we introduce two sets of matrices.
The one is \( \left( \F_A \right)_{BC} = f_{ABC} \) that defines the adjoint representation of the $U(N)$ Lie algebra.
The other one is \( \left( \D_A \right)_{BC} = d_{ABC} \). 

We recall two traced identities below, which hold for arbitrary matrices $A$, $B$, $C$, and $D$
\ba
{\rm Tr} 
\left(
\left[ A,B \right] \{ C,D \} + \left[ A,C \right] \{ B,D \} + \left[ A,D \right] \{ B,C \} 
\right) & = & 0,
\la{identity1} \\
{\rm Tr}
\left(
\left[ A,B \right] \left[ C,D \right] + \{ A,C \} \{ B,D \} - \{ A,D \} \{ B,C \} 
\right) & = & 0.
\la{identity2}
\ea
Then the application of (\ref{identity1}) and (\ref{identity2}) enables us to verify that
\ba
\left[ \D_i, \D_j \right] = \left[ \F_i, \D_j \right] & = & 0,
\la{ddfd} \\
\F_i \D_j + \F_j \D_i - d_{ijk} \F_k & = & 0,
\la{fdfd} \\
\F_i \F_j - \D_i \D_j + d_{ijk} \D_k & = & 0,
\la{ffdd}
\ea
where the subscripts $i,j,k$ = 1 to $N$. 
In addition, we have \( \left[ \F_i, \F_j \right] = 0 \) since $\F_i$ are the elements of Cartan subalgebra of the $U(N)$ Lie algebra in the adjoint representation. 
Using the explicit forms of the defining representation of $U(N)$, we can show that
\ba
\sum_i \left( \D_i \D_i \right)_{\alpha \beta}
& = & - \sum_i \left( \F_i \F_i \right)_{\alpha \beta} 
= \delta_{\alpha \beta},  
\la{sumffdd1} \\
\sum_i \left( \D_i \D_i \right)_{a b}
& = & - \sum_i \left( \F_i \F_i \right)_{a b}
= \delta_{a b},   
\la{sumffdd2} \\
\sum_i \left( \D_i \D_i \right)_{m n} 
& = & 2 \delta_{m n}.
\la{sumddij}
\ea
Furthermore, the combinations of (\ref{sumddij}) with (\ref{fdfd}) and (\ref{ffdd}) produce the matrix equations that are useful later. They are
\ba
d_{ijk} \F_j \D_k & = & \F_i,
\la{dfd} \\
d_{ijk} \D_j \D_k & = & \D_i,
\la{ddd} \\
d_{ijk} \F_j \F_k & = & - \D_i.
\la{dff}
\ea

Now we are in a position to generalize the method of Abelian decomposition for the $Sp(2N)$ Yang-Mills connection one-form
\be
A = A_\mu dx^\mu = 
\left( 
A_\mu^{0,\alpha} \T_{0,\alpha} + A_\mu^{I,a} \T_{I,a} + 
A_\mu^{I,i} \T_{I,i} \right) dx^\mu.
\la{Aym}
\ee
Based on the decomposition procedures presented in \cite{fad2,su}, we first conjugate the elements of Cartan subalgebra $\H_i$ (\ref{Hi}) by a generic element \( g \in Sp(2N) \) to generate $N$ Lie-algebra valued vector fields 
\be
m_i = g \H_i g^{-1}.
\la{mi}
\ee
The fields $m_i$ depend on $2N^2$ independent variables only, since they remain invariant if $g$ transforms by \( g \to gh \) for $h$ belongs to the maximal Abelian subgroup of $Sp(2N)$.

Next, using the fields $m_i$ (\ref{mi}), we parameterize the connection one-form $A$ (\ref{Aym}) as follows,
\be
A = C^i m_i + \frac{1}{i} \left[ dm_i, m_i \right] + ({\rm covariant~part}),
\la{Aoneform}
\ee
where \( dm_i = \partial_\mu m_i dx^\mu \) and $C^i$ are $U(1)$ connection one-forms.
In (\ref{Aoneform}), the first two terms on the right-hand-side are the Cho connection \cite{cho2}, which preserve the full non-Abelian gauge characteristics.
The third term on the right-hand-side, the (covariant part), by construction is orthogonal to the fields $m_i$ and transforms covariantly under gauge transformations. 

It turns out that (\ref{Aoneform}) can be represented in a manifestly gauge equivalent expression if we introduce the Maurer-Cartan one-form
\ba
R & = & g^{-1}dg, \non \\
  & = & R^{0,\alpha} \T_{0,\alpha} + R^{I,a} \T_{I,a} + R^{I,i} \T_{I,i} ,
\la{MConeform}
\ea
where \( g \in Sp(2N) \). Then, use (\ref{mi}) and (\ref{MConeform}) to rewrite the terms in (\ref{Aoneform}) 
\ba
dm_i 
& = & 
g \left[ R, \H_i \right] g^{-1}, 
\non \\
& = & 
\frac{i}{\sqrt{2}} g
\Bigg[ 
R^{3,a} \left( \F_i \right)_{a \alpha} \T_{0,\alpha} +
R^{0,\alpha} \left( \F_i \right)_{\alpha a} \T_{3,a} 
\non \\
& + & \epsilon_{I3J} 
\Big( R^{I,a} \left( \D_i \right)_{a b} \T_{J,b} +
R^{I,m} \left( \D_i \right)_{m n} \T_{J,n} \Big) 
\Bigg] g^{-1}, 
\la{dmi}
\ea 
where $\epsilon_{IJK}$ is the Levi-Civita tensor and 
\ba
\left[ dm_i, m_i \right] 
& = & 
dg g^{-1} - g \left( R^{3,i} \H_i 
              + \frac{1}{2} \left( R^{0,\alpha} \T_{0,\alpha} + R^{I,a} \T_{I,a} \right)
              \right) g^{-1}.
\ea
Consequently, the connection one-form (\ref{Aoneform}) becomes
\ba
A 
& = &
g {\tilde A} g^{-1} + \frac{1}{i} dg g^{-1},
\la{tildeA} \\
& = &
g \left[ 
\left( C^i - \frac{1}{i} R^{3,i} \right) \H_i
- \frac{1}{2i} \left( R^{0,\alpha} \T_{0,\alpha} + R^{I,a} \T_{I,a} \right) 
+ ({\rm C.P.}) \right] g^{-1}
+ \frac{1}{i} dgg^{-1},
\la{Aequivalent}
\ea
where \( ({\rm C.P.}) = g^{-1} ({\rm covariant~part}) g \).

Lastly, the local basis of the (C.P.) space in (\ref{Aequivalent}) can be determined by the gauge covariant Lie-algebra valued one-forms that are constructed by the modified adjoint action
\be
{\tilde\delta}^i v = \left[ v, \H_i \right],
\la{madjointaction}
\ee
where $v$ is an arbitrary Lie-algebra valued element.
It should be stressed that the (C.P.) space in (\ref{Aequivalent}) is coincident with the orbit \( Sp(2N)/U(1)^N \).
That is, the total number of independent variables of the Lie-algebra valued one-forms is $2N^2$.
This can be easily deduced by differentiating the degrees of freedom carried between the original $Sp(2N)$ Yang-Mills connection (\ref{Aym}) and the Cho connection in (\ref{Aoneform}). 

To find the set of the Lie-algebra valued one-forms in the (C.P.) space, we notice that the one-forms \( x_i \equiv \frac{1}{i} \left[ R, \H_i \right] \) (\ref{dmi}) are orthogonal to $\H_k$, since \( {\rm Tr} \left( x_i H_k \right) = 0 \).
Hence, the one-forms $x_i$ determine a part of the basis states of the (C.P.) space.  
Next, we apply the modified adjoint action (\ref{madjointaction}) on $x_i$ to obtain another one-forms $z_{ij}$,
\ba
z_{ij} & \equiv & {\tilde\delta}^j x_i = \left[ x_i, \H_j  \right],
\la{dxi} \\
& = & 
\frac{i}{2}
\Bigg[ 
R^{0,\alpha} \left( \F_i \F_j \right)_{\alpha \beta} \T_{0,\beta} +
R^{3,a} \left( \F_i \F_j \right)_{a b} \T_{3,b} 
\non \\
& - & {\hat\delta}_{IJ} 
\Big( R^{I,a} \left( \D_i \D_j \right)_{a b} \T_{J,b} +
R^{I,m} \left( \D_i \D_j \right)_{m n} \T_{J,n} \Big)
\Bigg], 
\la{zij}
\ea
where \( {\hat\delta}_{IJ} = \delta_{IJ} - \delta_{I3}\delta_{J3} \).

The one-forms $z_{ij}$ are also orthogonal to $\H_k$ and determine another part of the basis states of the (C.P.) space.
In this way, we find the complete set of the Lie-algebra valued one-forms, that determine the entire basis states, by repeatedly using the modified adjoint action (\ref{madjointaction}). 
After some computations, we find
\ba
{\tilde\delta}^k z_{ij} & = & -\frac{1}{8} 
                      \left[ d_{ikl}d_{ljm} + d_{jkl}d_{lim} \right] x_m 
                      -\frac{1}{2}
                      \left[ d_{ijl} \, v_{lk} + d_{ikl} \, v_{lj} + d_{jkl} \, v_{li} 
                      \right], 
\la{dzij} \\
{\tilde\delta}^k v_{ij} & = & \frac{1}{4}
                      \left[ d_{kil}d_{ljm} + d_{jil}d_{lkm} \right] y_m
                      -\frac{1}{4} 
                      \left[ d_{kil} \, z_{lj} + d_{jil} \, z_{lk} + d_{kjl} \, u_{li}
                      \right], 
\la{dvij} \\
{\tilde\delta}^j y_{i} & = & -\frac{1}{2} v_{ij},
\la{dYi} \\
{\tilde\delta}^k u_{ij} & = & -\frac{1}{8} 
                      \left[ d_{ikl}d_{ljm} + d_{jkl}d_{lim} \right] x_m 
                      -\frac{1}{2}
                      \left[ d_{ijl} \, v_{lk} + d_{ikl} \, v_{jl} + d_{jkl} \, v_{il} 
                      \right], 
\la{duij}
\ea
where the explicit expressions of the one-forms \( (v_{ij}, y_i, u_{ij}) \) are given below in terms of the coefficients of the Maurer-Cartan one-form (\ref{MConeform})
\ba
v_{ij}  
& = & 
\frac{i^2}{2\sqrt{2}}
\Bigg[ 
R^{3,a} \left( \D_i \F_j \right)_{a \alpha} \T_{0,\alpha} +
R^{0,\alpha} \left( \D_i \F_j \right)_{\alpha a} \T_{3,a} 
\non \\
& + & \epsilon_{I3J}
\Big( R^{I,a} \left( \D_i \D_j \right)_{a b} \T_{J,b} +
R^{I,m} \left( \D_i \D_j \right)_{m n} \T_{J,n} \Big)
\Bigg],
\la{vij} \\
y_i 
& = & 
\frac{i^3}{4}
\Bigg[ 
R^{0,\alpha} \left( \D_i \right)_{\alpha \beta} \T_{0,\beta} +
R^{3,a} \left( \D_i \right)_{a b} \T_{3,b} 
\non \\
& + & {\hat\delta}_{IJ}  
\Big( R^{I,a} \left( \D_i \right)_{a b} \T_{J,b} +
R^{I,m} \left( \D_i \right)_{m n} \T_{J,n} \Big) 
\Bigg], 
\la{yi} \\
u_{ij} 
& = & 
\frac{i^3}{2}
\Bigg[ 
R^{0,\alpha} \left( \D_i \D_j \right)_{\alpha \beta} \T_{0,\beta} +
R^{3,a} \left( \D_i \D_j \right)_{a b} \T_{3,b} 
\non \\
& + & {\hat\delta}_{IJ} 
\Big( R^{I,a} \left( \D_i \D_j \right)_{a b} \T_{J,b} +
R^{I,m} \left( \D_i \D_j \right)_{m n} \T_{J,n} \Big)
\Bigg]. 
\la{uij}
\ea

Presumably, we get five subsets of Lie-algebra valued one-forms \( (x_i, z_{ij}, v_{ij}, y_i, u_{ij}) \) in total, which form a closed algebra under the modified adjoint action (\ref{madjointaction}). 
These one-forms \( (x_i, z_{ij}, v_{ij}, y_i, u_{ij}) \) possess definite properties under $SO(N)$ symmetries.
The one-forms $x_i$ and $y_i$ yield the $SO(N)$ vector representations, $z_{ij}$ and $u_{ij}$ the $SO(N)$ symmetric tensor representations, and $v_{ij}$ the $SO(N)$ rank-two tensor representation.
However, the results of (\ref{dfd}), (\ref{ddd}) and (\ref{dff}) assert the following equalities among these one-forms
\ba
x_i = -2 d_{ijk} v_{jk},
\la{xv} \\ 
y_i = 2 d_{ijk} z_{jk} = 2 d_{ijk} u_{jk}.
\la{yzu}
\ea
It implies that the one-forms $x_i$ and $y_i$ are not independent at all and that the one-forms $u_{ij}$ satisfy $N$ constraint equations.
Accordingly, we can count the number of independent components carried by those of the relevant one-forms $z_{ij}$, $v_{ij}$, and $u_{ij}$.  
The dimension of the symmetric tensor $z_{ij}$ is $\frac{1}{2}N(N+1)$ and the dimension of the second rank tensor $v_{ij}$ is $N^2$. 
After taking the constraint (\ref{yzu}) into account, the dimension of the other symmetric tensor $u_{ij}$ is \( \frac{1}{2}N(N+1)-N \).
The sum of these three numbers is $2N^2$, which as expect matches the dimension of the space $Sp(2N)/U(1)^N$.
 
In order to proceed the decomposition of the $Sp(2N)$ connection (\ref{Aequivalent}), we need appropriate dual variables that appear as coefficients to the one-forms \( (z_{ij}, v_{ij}, u_{ij}) \).
We denote them by \( (\phi^{ij}, \psi^{ij}, \sigma^{ij}) \), respectively.
We observe that the Yang-Mills connection $A$ in (\ref{Aym}) is an $Sp(2N)$ Lie-algebra valued one-form and transforms in the scalar representation of the $SO(N)$ symmetry.
So the dual variables must be zero-forms and transform in the same $SO(N)$ representations as the associated one-forms in order to form invariant combinations.

After including the dual variables \( (\phi^{ij}, \psi^{ij}, \sigma^{ij}) \), the connection one-form ${\tilde A}$ (\ref{tildeA}) takes the form 
\be
{\tilde A} = 
\left( C^i - \frac{1}{i} R^{3,i} \right) \H_i -
\frac{1}{2i} \left( R^{0,\alpha} \T_{0,\alpha} + R^{I,a} \T_{I,a} \right) +  
\phi^{ij} z_{ij} + \psi^{ij} v_{ij} + \sigma^{ij} u_{ij}.
\la{Adecomposed}
\ee
Due to the property of gauge symmetry (\ref{tildeA}) and (\ref{Aequivalent}), this connection one-form (\ref{Adecomposed}) can also be represented in a gauge equivalent expression in terms of the fields $m_i$ (\ref{mi}) 
\be
A = C^i m_i + \left( \delta_{ij} + \phi^{ij} \right) Z_{ij} + 
    \psi^{ij} V_{ij} + \sigma^{ij} U_{ij},
\la{Aconnection}
\ee
where \( \left( Z_{ij}, V_{ij}, U_{ij} \right) \) = \( g \left( z_{ij}, v_{ij}, u_{ij} \right) g^{-1} \).
It is apparent that \( (m_i, Z_{ij}, V_{ij}, U_{ij}) \) yields a complete set of basis states for the $Sp(2N)$ Lie algebra.
We therefore conclude that the decomposition of the four-dimensional $Sp(2N)$ connection (\ref{Aconnection}) contains the correct number of independent variables, which are appropriate for describing the theory in the infrared limit.

This work was supported in part by Taiwan's National Science Council Grant No. 89-2112-M-194-022.

\end{document}